\documentstyle[aps,prl,preprint,floats,epsfig]{revtex}

\begin{document}

\preprint{\tighten\vbox{\hbox{\hfil CLNS 99/1620}
                        \hbox{\hfil CLEO 99-4}
}}

\title{Evidence of New States Decaying into $\Xi_c^{*}\pi$}

\author{CLEO Collaboration}
\date{\today}

\maketitle
\tighten

\begin{abstract} 
Using data recorded by the CLEO II detector at CESR, we report
evidence for
two new charmed baryons, one decaying into $\Xi_c^+\pi^+\pi^-$ via an
intermediate $\Xi_c^{*0}$, and its isospin partner decaying
into $\Xi_c^0\pi^+\pi^-$ via an intermediate $\Xi_c^{*+}$.
We measure the mass differences of the two states to be
$M(\Xi_c^+\pi^+\pi^-)-M(\Xi_c^+)=$
$348.6\pm0.6\pm1.0$ MeV, and  $M(\Xi_c^0\pi^+\pi^-)-M(\Xi_c^0)=$
$347.2\pm0.7\pm2.0$ MeV. We interpret these new
states as the $J^P = {3 \over{2} }^-\ $ $\Xi_{c1}$ particles,
the charmed-strange analogues of the $\Lambda_{c1}^+(2625)$.

\end{abstract}
\newpage

{
\renewcommand{\thefootnote}{\fnsymbol{footnote}}

 
\begin{center}
J.~P.~Alexander,$^{1}$ R.~Baker,$^{1}$ C.~Bebek,$^{1}$
B.~E.~Berger,$^{1}$ K.~Berkelman,$^{1}$ V.~Boisvert,$^{1}$
D.~G.~Cassel,$^{1}$ D.~S.~Crowcroft,$^{1}$ M.~Dickson,$^{1}$
S.~von~Dombrowski,$^{1}$
% ...
\end{center}
 
\small
\begin{center}
$^{1}${Cornell University, Ithaca, New York 14853}\\
$^{2}${University of Florida, Gainesville, Florida 32611}\\
% ...
\end{center}

\setcounter{footnote}{0}
}
\newpage

In recent years there has been great progress in charmed
baryon spectroscopy. 
Three experiments\cite{LCS} 
have now seen a doublet of particles decaying into 
$\Lambda_c^+\pi^+\pi^-$, and the consensus is that these states are the 
lowest lying orbitally excited states of the $\Lambda_c^+$. The quark model 
picture of these excited $\Lambda_c^+$ baryons 
is that they consist of a light diquark
which has one unit of orbital angular momentum with respect to the 
heavy (charmed)
quark, leading to a $J^P={1\over{2}}^-,{3\over{2}}^-$ doublet. They
are now commonly referred to as the $\Lambda_{c1}^+$ particles\cite{CHO}, 
where the 
numerical subscript refers to the total angular momentum 
of the light degrees of freedom.  
Clearly similar orbital excitations must exist in the $\Xi_c^+$ sector. 
Using data from the CLEO II detector, we present the first evidence
of two new states, one  
decaying into $\Xi_c^+\pi^+\pi^-$ via an intermediate
$\Xi_c^{*0}$, and the other decaying into $\Xi_c^0\pi^+\pi^-$
via an intermediate $\Xi_c^{*+}$.
We identify these states as the
$J^P={3\over{2}}^-$ $\Xi_{c1}$ isospin doublet.  
Such states correspond to $csq$ 
quark combinations where $q$ is a $u$ or $d$ quark, 
the $q$ and $s$ spins are antiparallel, and
the $qs$ diquark has orbital 
angular momentum $L=1$ with respect to the charmed quark. 
Preliminary versions of this analysis were presented 
elsewhere \cite{CONF,CONF2}. The analysis presented here
includes mass dependent fitting of the particle trajectories
taking into account energy loss throughout the detector, improved 
secondary and tertiary vertex detection, and an increased number
of $\Xi_c$ decay modes used for $\Xi_c^*$ reconstruction.

The data presented here 
were taken by the CLEO II detector\cite{KUB} operating at the Cornell 
Electron Storage Ring.
The sample used in this analysis corresponds to
an integrated luminosity of 4.8 $fb^{-1}$ from data
taken on the $\Upsilon(4S)$ 
resonance and in the continuum at energies just above and below 
the $\Upsilon(4S)$.
We detected charged tracks with a cylindrical drift chamber system inside
a solenoidal magnet. Photons were detected using an electromagnetic
calorimeter consisting of 7800 cesium iodide crystals.

We first obtain large samples of reconstructed $\Xi_c^+$ and $\Xi_c^0$ 
particles, using 
their decays into $\Lambda$, $\Xi^-$, $\Omega^-$ and $\Xi^0$ hyperons
as well as $K$'s, $\pi$'s and protons\footnote
{Charge conjugate states are implied throughout.}. 
The analysis chain for reconstructing these particles follows closely 
that presented in our previous publications \cite{XIC1}.

We fitted the invariant mass distributions for each decay mode to a sum
of a Gaussian signal function and a second order 
polynomial background. The yields from all the decay modes
are summarized in Table 1. We note that this is the first observation of the 
decay modes $\Xi_c^+\to\Lambda\overline{K}^0\pi^+$ and 
$\Xi_c^0\to\Lambda K^-\pi^+$. 
$\Xi_c$ candidates were defined as those combinations within $2\sigma$ of the known
mass of the $\Xi_c^+$ or $\Xi_c^0$, where $\sigma$ is the detector resolution 
measured mode-by-mode by a Monte Carlo simulation program.
To illustrate the good statistics and
signal to noise ratio of the $\Xi_c$ signals, 
we have placed a cut 
$x_p>0.5$, where $x_p=p/p_{max}$, $p$ is the momentum 
of the charmed baryon, $p_{max}=\sqrt{E^2_{beam}-M{^2}},$ and
$M$ is the calculated $\Xi_c$ mass, and present the results for the various decay
modes in Table 1.
In the final analysis we prefer to apply 
an $x_p$ cut only on the $\Xi_c\pi^-\pi^+$ combinations.

\begin{table}[htb]
\caption{Measured yield for each sub-mode}
\begin{tabular}{cc}
$\Xi_c$ Decay Mode&$\Xi_c$ Yield ($x_p>0.5$)\\
\hline
$\Xi^-\pi^+\pi^+            $        & $369\pm24$        \\
$\Xi^0\pi^+\pi^0            $        & $231\pm30$        \\
$\Lambda\overline{ K^0}\pi^+$        & $61\pm13$         \\
\hline
$\Xi^-\pi^+\pi^0$                    & $130\pm19$        \\
$\Omega K^+$                         & $37\pm7$          \\
$\Xi^-\pi^+$                         & $230\pm18$        \\
$\Xi^0\pi^+\pi^-$                    & $103\pm22$        \\
$\Lambda K^-\pi^+$                   & $86\pm14$         \\
$\Lambda{\overline{K^0}}$            & $33\pm10$         \\
\end{tabular}
\end{table}

The $\Xi_c$ candidates defined above were then 
combined with each remaining charged track in the event 
and the mass differences $\Delta M$=
$M(\Xi_c^+\pi^-)-M(\Xi_c^+)$ and $M(\Xi_c^0\pi^+)-M(\Xi_c^0)$ were calculated. 
We consider those combinations within 5 MeV of the previously measured 
$\Xi_c^*$ peaks found in these plots\cite{XIC1} as $\Xi_c^*$ candidates.

We then combine these $\Xi_c^{*}$ candidates with one more
correctly 
charged track in the event and plot $M(\Xi_c\pi^-\pi^+)-M(\Xi_c)$ for both
the $\Xi_c^+$ (Figure 1a) and the $\Xi_c^0$ (Figure 1b),
with a requirement of $x_p>0.6$ on the final combination. 
We prefer to present the data as a dipion mass difference rather than 
$M(\Xi_c^{*}\pi)-M(\Xi_c^{*})$ 
because the latter measurement is complicated by the intrinsic width of the 
$\Xi_c^{*}$, which is not well known.
In both figures there is a peak at around 348 MeV. 
We fit these two peaks to sums of 
Gaussians of fixed width ($\sigma=1.8$ MeV,  
found from simulated events), and a
polynomial background function. For the charged case, 
we find a signal of $19.7\pm4.5$
events at a $\Delta M$ of $348.6\pm0.6$ MeV. 
For the neutral case, we find an excess of $9.5\pm3.2$ events 
at $\Delta M $ of $347.2\pm0.7$ MeV. 
Both peaks are satisfactorily fit using this fitting function,
however, to investigate the natural widths of these orbitally excited
states,
we have also fit to a Breit-Wigner function 
convoluted with a Gaussian resolution function.
This gives limits to the natural widths of the states of
$\Gamma < 3.5 $ MeV and $\Gamma < 6.5 $ MeV respectively, each at the 90\% 
confidence level. We estimate the systematic uncertainty on the measured
mass differences to be 1 MeV and 2 MeV respectively. This estimate
takes into account the spread of results obtained using different 
fitting functions as well as uncertainties in the momentum measurements.
The systematic uncertainty in the neutral case is 
large as this measurement is particularly sensitive to the
choice of fitting function.

In order to check that all the $\Xi_{c1}$ decays proceed via an 
intermediate $\Xi_c^*$, we release the cuts on $M(\Xi_c\pi)-M(\Xi_c)$, 
select combinations within
5 MeV of our final signal peaks and plot $M(\Xi_c\pi)-M(\Xi_c)$. 
Both plots (Figures 2a and b) show signals which were fit to a Breit-Wigner convoluted with a 
Gaussian resolution function, plus a flat background. The masses and widths 
for the $\Xi_c^*$ particles found in this way are consistent with our
previously published results. It is clear that the data is consistent with
all the $\Xi_{c1}$ decays proceeding via an intermediate $\Xi_c^*$.

Although the statistics are very limited, they are sufficient to do a rough 
investigation of the momentum spectrum with which the new particles are produced. 
We add the two isospin states together as we would expect them to have very
similar momentum distributions.
We relax the $x_p$ cut from 0.6 to 0.5, and fit the dipion mass difference plots
(Figure 2) in bins of $x_p$. The fit uses a fixed width derived from the Monte-Carlo
study, with the mass fixed at the value found for $x_p>0.6$. 
The yields in each 
bin of $x_p$ were corrected for the detector efficiency, and the resulting
 $x_p$ distribution shown in Figure 5. 
The fit to this spectrum is of the functional form
due to Peterson $et\ al.$\cite{PETER}. 
The fitted parameter $\epsilon_r$ is measured to 
be $\epsilon_r=0.07^{+0.03}_{-0.02}$. This value is very similar
to that found for the $\Lambda_{c1}^+$ spectrum\cite{LCS}, 
and harder than those found for charmed baryons with no orbital 
angular momentum.

There has been little theoretical work in recent years on the spectroscopy of 
orbitally excited $\Xi_c$ states. However, the models\cite{MI} 
that do exist predict that the
excitation energy of the first orbitally excited doublet should be 
similar to the 
analogous value in the $\Lambda_c^+$ case 
(308 and 342 MeV for the two states). Furthermore
the decay patterns of the $\Xi_{c1}$ states should be 
closely analogous to those of 
the $\Lambda_{c1}^+$. 
The preferred decay of the $J^P={3\over{2}}^-$ $\Xi_{c1}$ 
should be to $\Xi_c^*\pi$ because the 
spin-parity of the baryons allows this decay to
proceed via an $S$-wave decay. Decays to $\Xi_c^{\prime}$ would have to 
proceed via a D-wave and would therefore be suppressed. 
In the case of the $J^P={1\over{2}}^-$
$\Xi_{c1}$ the situation is reversed. It is natural therefore to 
expect a particle 
found by its decay to $\Xi_c^*\pi$ to have $J^P={3\over{2}}^-$. 
When the total spin and
parity of the baryon is considered, decays directly to the 
ground state with one transition $\pi$, 
not allowed for the $\Lambda_{c1}^+$ because of isospin conservation,
are allowed for the $\Xi_{c1}$ via a D-wave. Taking into account  
the large phase space available, such decays might be expected to be large. 
However, in the Heavy Quark Effective
Theory (HQET)\cite{IW}, where the angular momentum 
and parity of the light diquark degrees of freedom must be
considered separately from those of the heavy quark, such decays are forbidden.
Thus in the HQET picture, we would expect that the dominant decay of a 
$J^P={3\over{2}}^-$ $\Xi_{c1}$ would be to $\Xi_c^*\pi$, 
consistent with our observation. 

Following our first analysis of the $\Xi_{c1}^+$, there
have been two papers that include theoretical calculations of the expected 
$\Xi_{c1}$ widths. Yan and Pirjol\cite{PIRJ} calculate $2.37-15.00$ 
MeV, whereas 
Chiladze and Falk \cite{FALK} calculate $5.4^{+3.8}_{-2.5}$ MeV. 
Both calculations use the 
experimentally measured width of the $\Lambda_{c1}(2593)$ as input. Our results 
favour a natural width for each of the $\Xi_{c1}$ particles at the lower end 
of these predictions.
   
In conclusion, we present evidence for the production of two new states. 
The first
of these states
decay into $\Xi_c^{*0}\pi^+$ with measured mass given by 
$M(\Xi_c^+\pi^-\pi^+)-M(\Xi_c^+)=348.6\pm0.6\pm1.0$ MeV, and 
width, $\Gamma<3.5$ MeV at the 90\% confidence level.
The second state decays into $\Xi_c^{*+}\pi^-$ with a mass given by
$M(\Xi_c^0\pi^+\pi^-)-M(\Xi_c^0)$ = $347.2\pm0.7\pm2.0$ MeV, and width,
$\Gamma<6.5$ MeV at the 90\% confidence level. Although we
do not measure the spin or parity of these states, the observed 
decay modes, masses, 
and momentum distributions are all consistent with the new 
states being the $J^P={3\over{2}}^-$ $\Xi_{c1}^+$ and $\Xi_{c1}^0$ states, 
the charmed-strange
analogues of the $\Lambda_{c1}^+(2625)$.

\bigskip
\begin{figure}[tb]
\centering
\psfig{file=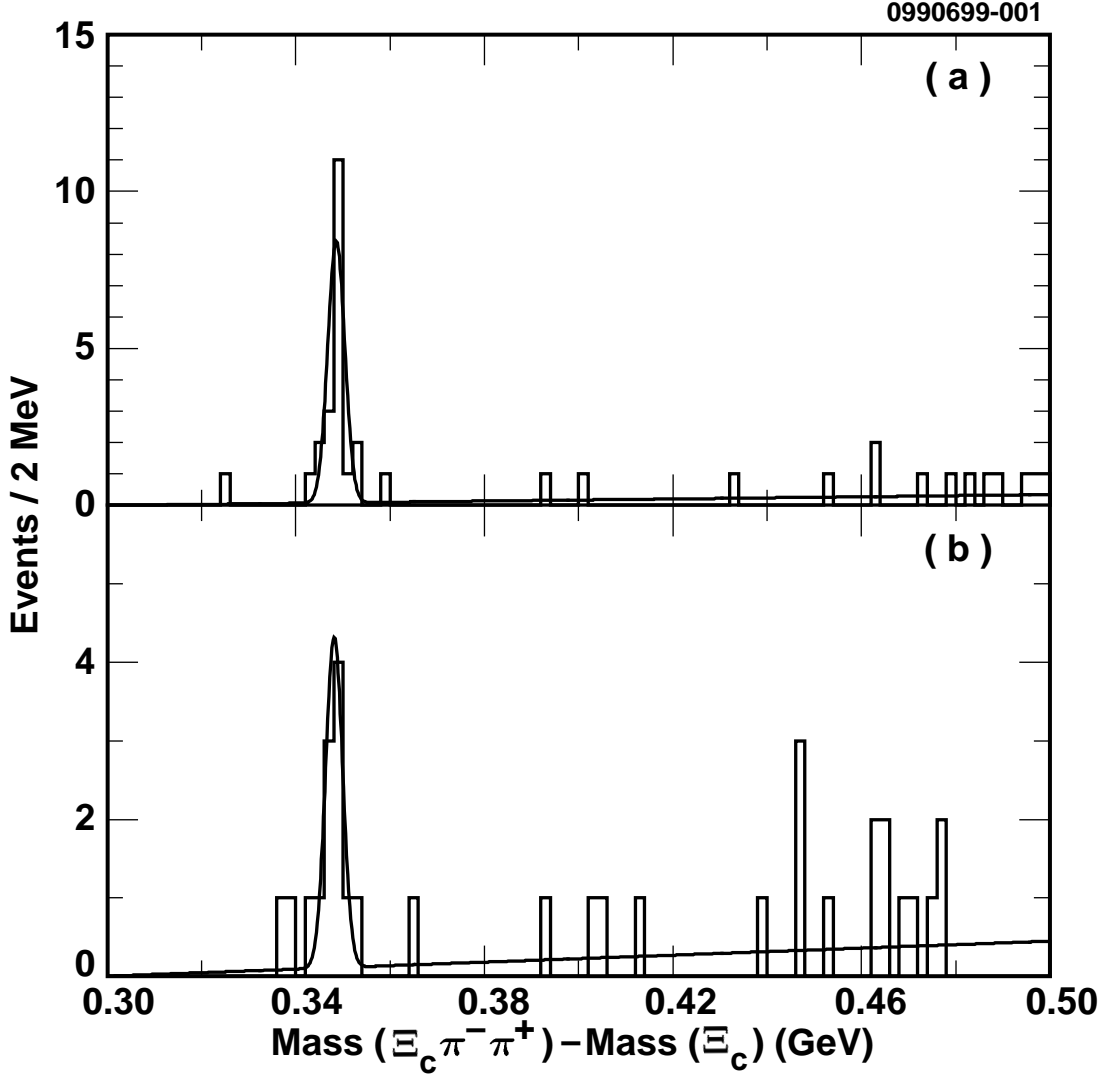}
\caption[]{(a)The mass difference $M(\Xi_c^+\pi^+\pi^-)-M(\Xi_c^+)$
for combinations that lie in the $\Xi_c^{*0}$ band.
and (b) the mass difference $M(\Xi_c^0\pi^+\pi^-)-M(\Xi_c^0)$ for combinations 
that lie in the $\Xi_c^{*+}$ band. In both cases an $x_p>0.6$ cut is applied.}
\end{figure}
\begin{figure}[tb]
\centering
\psfig{file=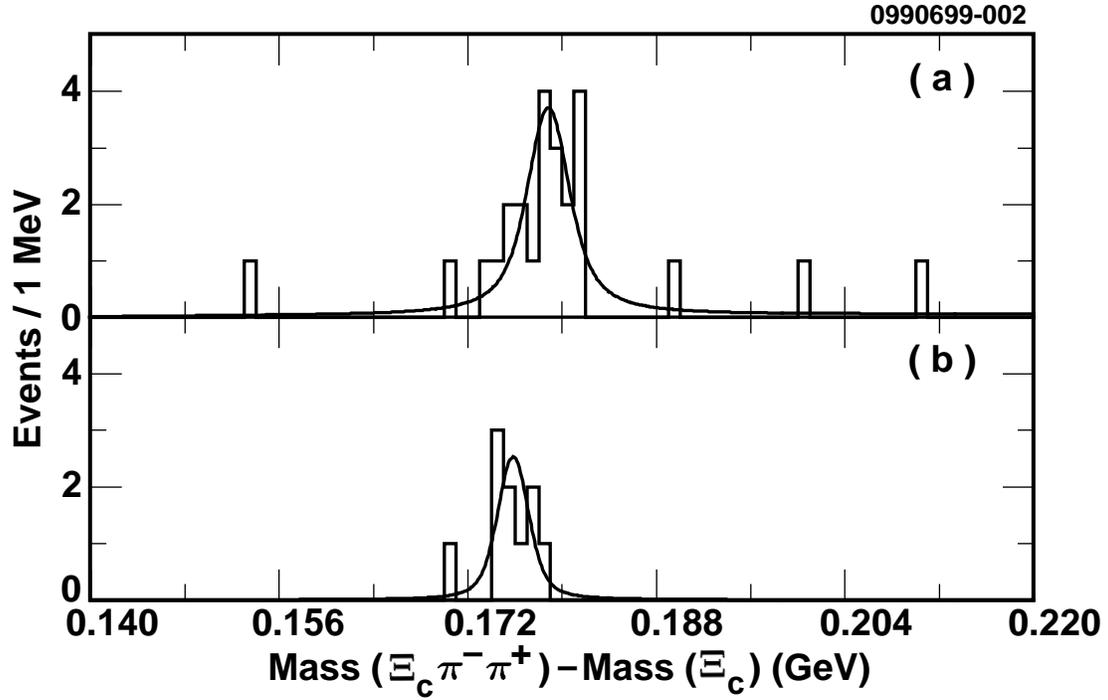}
\caption[]{After selecting combinations in the 
signals shown in Figure 1, we plot 
(a) $M(\Xi_c^+\pi^-)-M(\Xi_c^+)$ and (b) $M(\Xi_c^0\pi^+)-M(\Xi_c^0)$. 
Clear $\Xi_c^*$ peaks are seen 
in both plots.}
\end{figure}
\begin{figure}[tb]
\centering
\psfig{file=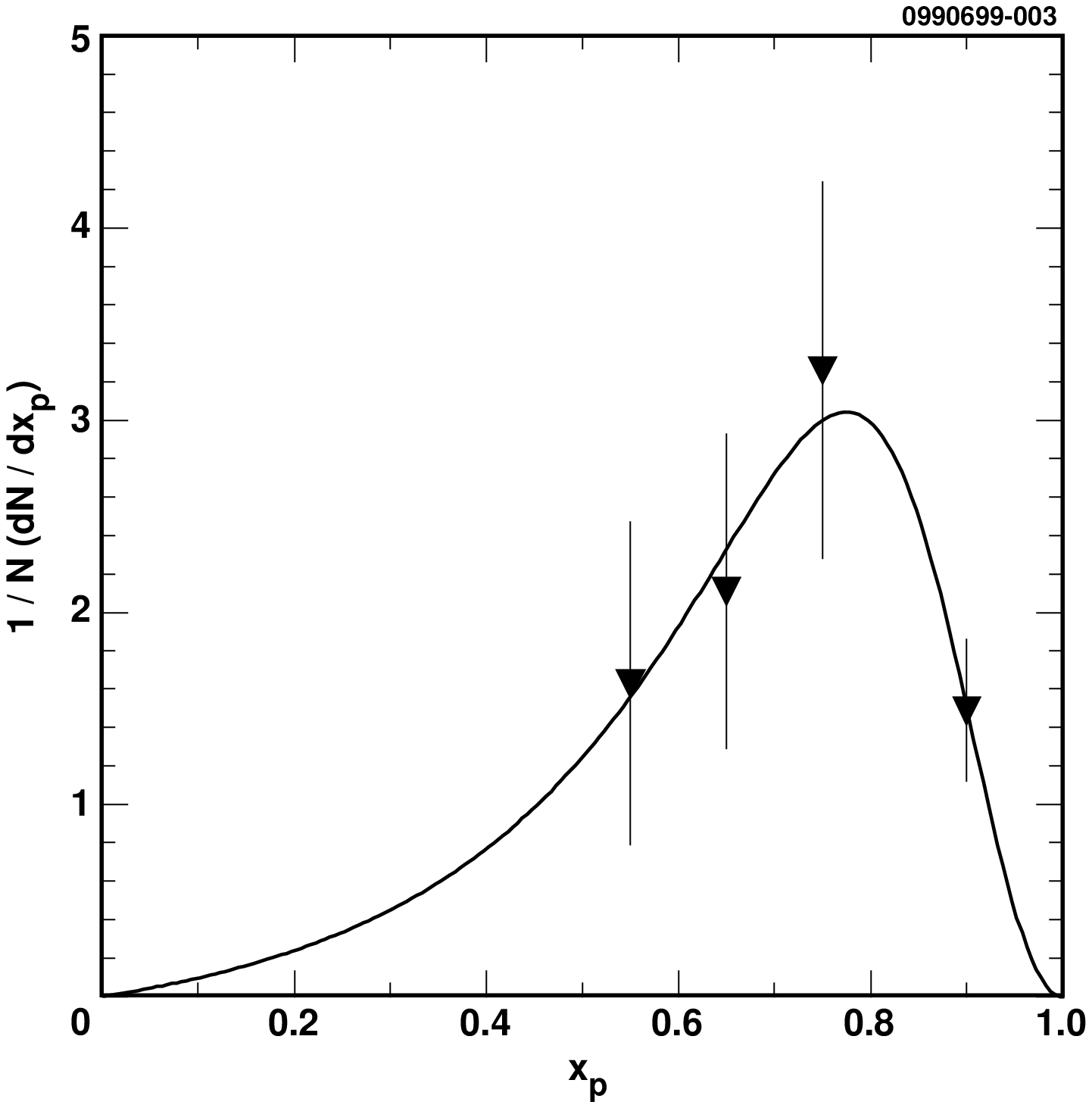}
\caption[]{Scaled momentum distribution of the new particles. The fit 
is of the form of the Peterson function.}
\end{figure}

We gratefully acknowledge the effort of the CESR staff in providing us with
excellent luminosity and running conditions.
This work was supported by 
the National Science Foundation,
the U.S. Department of Energy,
the Research Corporation,
the Natural Sciences and Engineering Research Council of Canada, 
the A.P. Sloan Foundation, 
the Swiss National Science Foundation, 
and the Alexander von Humboldt Stiftung.

{}

\end{document}